\documentclass[twocolumn,preprintnumbers,amsmath,amssymb]{revtex4}

\usepackage{graphicx}
\usepackage{dcolumn}
\usepackage{bm}

\begin{document}

\title{Quantifying Volume Changing Perturbations in a Wave Chaotic System}

\author{Biniyam Tesfaye Taddese\textsuperscript {2}}
\author{Gabriele Gradoni\textsuperscript {1}}
\author{Franco Moglie\textsuperscript {3}}
\author{Thomas M. Antonsen\textsuperscript {1, 2}}
\author{Edward Ott\textsuperscript {1, 2}}
\author{Steven M. Anlage\textsuperscript {1, 2}}
\affiliation{Department of Physics\textsuperscript {1}, University of Maryland, College Park, Maryland 20742-3285, USA.
Department of Electrical and Computer Engineering\textsuperscript {2}, University of Maryland, College Park, Maryland 20742-4111, USA.
Dipartimento di Ingegneria dell'Informazione\textsuperscript {3}, Universita Politecnica delle Marche, Ancona, Italy.
}

\date{\today}

\begin{abstract}
A sensor was developed to quantitatively measure perturbations which change the volume of a wave chaotic cavity while leaving its shape intact. The sensors work in the time domain by using either scattering fidelity of the transmitted signals or time reversal mirrors. The sensors were tested experimentally by inducing volume changing perturbations to a one cubic meter mixed chaotic and regular billiard system. Perturbations which caused a volume change that is as small as $54$ parts in a million were quantitatively measured. These results were obtained by using electromagnetic waves with a wavelength of about $5cm$, therefore, the sensor is sensitive to extreme sub-wavelength changes of the boundaries of a cavity. The experimental results were compared with Finite Difference Time Domain (FDTD) simulation results, and good agreement was found. Furthermore, the sensor was tested using a frequency domain approach on a numerical model of the star graph, which is a representative wave chaotic system. These results open up interesting applications such as: monitoring the spatial uniformity of the temperature of a homogeneous cavity during heating up / cooling down procedures, verifying the uniform displacement of a fluid inside a wave chaotic cavity by another fluid, etc.
\end{abstract}
\maketitle

\clearpage

\section{Introduction \label{sec:3-Introduction}}

Many sensors, such as SONAR, rely on ballistic wave propagation that provides only direct line of sight information. In that case, monitoring a cavity which has an irregular geometric shape, including hidden regions, may require installing multiple sensors throughout the cavity for a comprehensive coverage. However, most real world cavities have irregular shapes. This irregularity has the benefit of facilitating the creation of ray chaotic trajectories. The study of waves propagating inside these ray chaotic cavities, in the semi-classical limit, is called wave chaos \cite{Stockmann1999, Ott2002}. Wave chaos is essentially the manifestation of the underlying ray chaos on the properties of the waves whose wavelength is much smaller than the typical dimensions of the cavity. For instance, random matrix theory has been shown to describe the spectral statistics of quantum systems with chaotic classical counterparts \cite{Bohigas1984, Hemmady2005a, Hemmady2005, Hemmady2006a}. Ray chaos is characterized by sensitive dependence of ray trajectories to initial conditions. The effect of perturbations on waves propagating in such cavities was studied using the concept of the scattering fidelity \cite{Gorin2006}. Scattering fidelity is a normalized correlation between two cavity response signals as a function of time; the response signals are typically collected before and after a perturbation to the cavity. The scattering fidelity decay resulting from global (as opposed to local) perturbations which change all the boundary conditions of wave chaotic cavities has been studied \cite{Lobkis2003, Gorin2006a, Lobkis2008}. Global perturbations that change only one of the walls of the cavity have also been considered \cite{Schafer2005, Schafer2005a}. The scattering fidelity decay associated with global perturbations is either an exponential or Gaussian function of time. On the other hand, it has been shown that the scattering fidelity associated with a local perturbation has a slower algebraic decay \cite{Hohmann2008}. The fidelity decay induced by perturbing the cavity coupling has also been considered \cite{Kober2010}. On the other hand, the effect of a local boundary perturbation on a quantum mechanical system is studied using the Loschmidt echo concept\cite{Goussev2007, Kober2011}.

Despite the extensive research on scattering fidelity, a practical sensing application of the scattering fidelity concept has not been explored. In previous work, wave chaotic sensing techniques that allow a comprehensive spatial coverage using a single sensor were introduced \cite{Taddese2009, Taddese2010, Taddese2009a}. These techniques rely on the wave chaotic nature of most real world cavities. When a wave is broadcast into a cavity to probe it, the response signal consists of reflections that bounce from almost all parts of the cavity; this is due to the underlying spatial ergodicity of ray trajectories in ray chaotic cavities. Therefore, the response signal essentially "fingerprints" the cavity, and it enables the detection of changes to the cavity.

The wave chaotic sensing techniques developed in Refs.\cite{Taddese2009, Taddese2010, Taddese2009a} were not used to quantify any kind of perturbation. Local and global perturbations to the boundaries of the cavity, and perturbations to the medium of wave propagation within the cavity, were all shown to be detectable \cite{Taddese2010, Kober2011}. However, the quantification of a perturbation was not accomplished. On the other hand, a remarkably sensitive quantification of a perturbation which involved translation of a sub-wavelength object over sub-wavelength distances was successfully demonstrated \cite{Cohen2011}. However, the quantification was based on an empirical law that is specific to the system and perturbation at hand. This is because the effect of the perturbation on the dynamics of the waves propagating inside the wave chaotic cavity is not straightforward \cite{Cohen2011}. In this paper, we focus on a single class of perturbation whose effect can be theoretically predicted, and we propose two time domain techniques to measure that particular kind of perturbation in any cavity.

In this paper, we focus on quantifying volume changing perturbations (VCP) to a wave chaotic scattering system. Such systems have all degeneracies broken, and we shall further assume that they are time-reversal invariant. In general, a VCP changes the volume of a cavity, but it may slightly change its shape as well. A special kind of VCP is a volume changing and shape preserving perturbation (VCSPP).  In Sec.~\ref{sec:3-Theory}, the theoretical prediction of the effect of VCSPPs is discussed. Sec.~\ref{sec:3-Theory} proposes two time domain techniques to quantify VCPs. As in previous work, these techniques are based on the scattering fidelity and time reversal mirrors \cite{Taddese2009, Taddese2010, Taddese2009a}, whose experimental implementations are distinctly different. Sec.~\ref{sec:3-Testing} presents the experimental test of these two VCSPP sensing techniques, along with a head-to-head numerical validation. The experimental test is carried out inside a mixed chaotic and regular billiard system using electromagnetic waves. Sec.~\ref{sec:3-FreqDomain} provides a test of the sensing techniques in a numerical model of the star graph, which is a quasi-1D wave chaotic system. Sec.~\ref{sec:3-FreqDomain} also shows the relative merits of approaching the problem in the frequency domain. Sec.~\ref{sec:3-discussion} discusses practical applications of the VCSPP sensor, and Sec.~\ref{sec:3-conclusion} provides a conclusion.

\section{Theory and Approach \label{sec:3-Theory}}

\begin{figure}
\begin{center}
\includegraphics[width=3in]{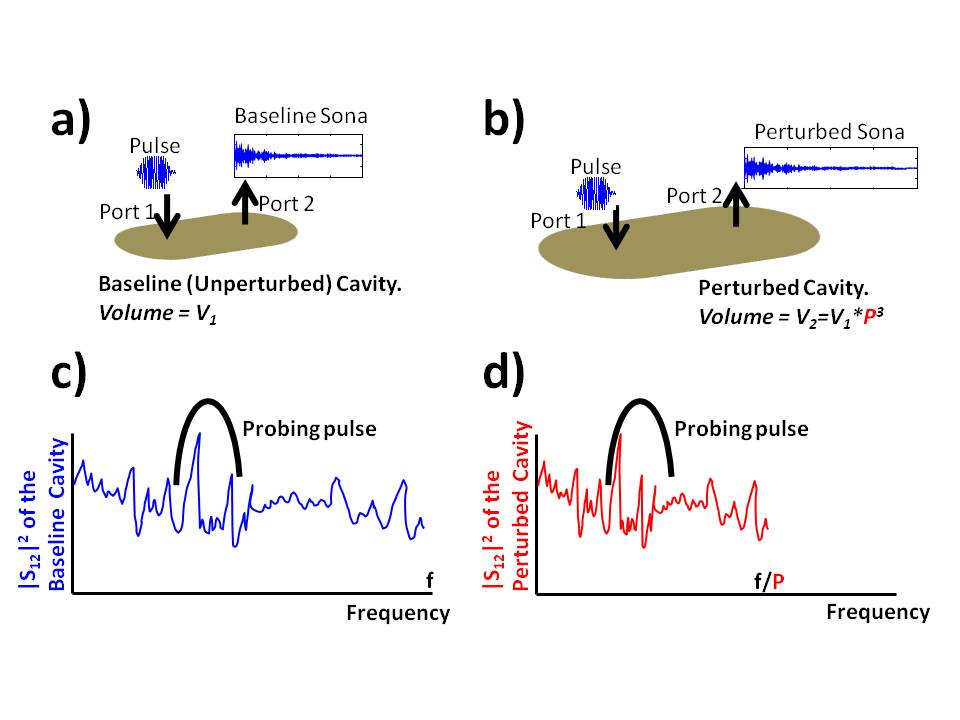}
\end{center}
\begin{quote}
\caption{Schematic illustrating Volume Changing \& Shape Preserving Perturbations (VCSPP). (a) Sona is collected from a baseline cavity of volume $V_{1}$. (b) Sona is collected from a perturbed cavity of volume $V_{2}$. (c) Pulse exciting the resonances of the baseline system. (d) The same pulse exciting the perturbed resonances of the perturbed system. \label{fig:vcpFig1}}
\end{quote}
\end{figure}

Consider a generic wave chaotic cavity with volume $V_{1}$, which is considered as a baseline (i.e. reference or unperturbed) system (see Fig.~\ref{fig:vcpFig1}(a)). The schematics in Fig.~\ref{fig:vcpFig1}(a)\&(b) illustrate the cavity as a stadium billiard, but the cavity is considered generic throughout Sec.~\ref{sec:3-Theory}. Suppose that the baseline cavity is perturbed such that each of its three length dimensions increase by a factor of $P$. This amounts to a VCSPP, by a factor of $P^{3}$; the perturbed cavity has a volume of $V_{2}=P^{3}V_{1}$ (see Fig.~\ref{fig:vcpFig1}(b)). Fig.~\ref{fig:vcpFig1}(a)\&(b) show a brief pulse being broadcast into the cavity. The response signal to the pulse is called the sona. The typical durations of the pulse and the sona are $1ns$ and $10\mu s$ respectively. The sona from the baseline cavity (which is referred to as baseline sona) and the sona from the perturbed cavity (which is referred to as perturbed sona) are expected to be related, under certain conditions which are discussed in Sec.~\ref{sec:3-LimitationTD}. For instance, if $P>1$, a signal feature in the perturbed sona is expected to be delayed by a factor of $P$ compared to its appearance in the baseline sona. The sensor is designed to enable the measurement of the value of $P$, which effectively quantifies the VCSPP, by using the theoretically predicted effects of the VCSPP on the dynamics of the waves. Another practically useful capability of the sensor is to check if the perturbation is indeed a VCSPP, and not just merely a VCP.

Next, consider the scattering parameters of the cavities as a function of frequency. The $|S_{12}|^{2}$ as a function of frequency of the baseline and the perturbed $2$-port cavity are schematically shown in Figs.~\ref{fig:vcpFig1}(c)\&(d). Let us assume that the antennas coupling energy into the cavities have a negligible frequency dependence. Then, we expect a precise mathematical relationship between the scattering parameters of the baseline and perturbed cavities as a function of frequency. In particular, if $P>1$, the baseline spectrum can be obtained by stretching out the perturbed spectrum by a factor of $P$ along its frequency axis. This is precisely the prediction about the effect of VCSPPs on the dynamics of the waves. This prediction can be used to measure the perturbation in the frequency domain. As opposed to the resource intensive frequency domain sensing, it is practically preferable to use a time domain interrogation of the baseline and the perturbed cavity by measuring the sonas. However, it is useful to look at the problem in the frequency domain to understand the limitations of the time domain approach discussed in Sec.~~\ref{sec:3-LimitationTD}.

There are two classes of time domain sensing techniques that can be used to quantify VCSPPs. The first technique relies on the scattering fidelity \cite{Gorin2006}. Consider two sonas $X$ and $Y$, which are real voltage versus time signals with zero mean values. The scattering fidelity ($SF$) of $X$ and $Y$ is simply their Pearson's correlation as a function of time, $t$ \cite{Gorin2006};
\begin{equation}
SF(t)=\frac{ \sum_{m=t}^{m=t+\Delta t}X[m]Y[m] }{ \sqrt { \sum_{m=t}^{m=t+\Delta t}X[m]^{2} \sum_{m=t}^{m=t+\Delta t}Y[m]^{2} } }
\label{eqn:sf}
\end{equation}
where $\Delta t$ is typically chosen to be the time it takes the waves to traverse the cavity, at the very least, once (i.e. in order of magnitude of the ballistic flight time). The $SF(t)$ of two sonas can take real values ranging from $1$ (i.e. perfect correlation at time $t$) to $-1$ (i.e. perfect anti-correlation at time $t$). If $SF(t)$ is $0$, then the sonas are not correlated at time $t$. The $SF$ of the baseline and the perturbed sonas is not expected to stay close to $1$ throughout time. However, the $SF$ of the perturbed sona and the baseline sona whose time axis is scaled using the optimum stretching/squeezing factor is expected to stay close to $1$ throughout time. The optimum stretching/squeezing factor is expected to be equal to $P$, which is also related to the magnitude of the perturbation.

The second technique to quantify VCSPPs utilizes classical time reversal mirrors \cite{Fink1996}. To see the operation of a time reversal mirror, consider a two port cavity. Suppose that a pulse is broadcast into the baseline cavity through port 1, and a baseline sona is recorded through port 2. If the baseline sona is time reversed and broadcast back into the cavity through port 2, a time-reversed version of the original pulse reconstructs at port 1. The reconstructed pulse approximates the time reversed version of the original pulse broadcast into the cavity. However, if the time reversed baseline sona is broadcast into a perturbed cavity, then the reconstructed pulse will more poorly approximate the time reversed version of the original pulse. The time axis of the baseline sona needs to be scaled using the optimum factor before it is time reversed and broadcast into the perturbed cavity; this is assuming that the perturbation is VCSPP. The optimum stretching/squeezing factor is expected to result in a reconstructed pulse that best approximates the original pulse. Once again, the optimum stretching/squeezing factor is expected to be $P$.

Time-reversal mirrors have found a wide range of practical applications such as crack imaging in solids \cite{Ulrich2007, Ulrich2009}, and improved acoustic communication in air \cite{Candy2006}, among other things \cite{Fink1996, Lerosey2005}. Time-reversal mirrors have been shown to benefit from the cavity's underlying ray chaos, which is prevalent in most real world cavities \cite{Calvo2008}. Recently, it was proposed that time reversal mirrors could also be applied to quantum systems \cite{Pastawski2007}. The robustness of time-reversal mirrors in a scattering medium undergoing perturbations has also been studied \cite{Tourin2001}.
\section{Testing the Quantitative Sensor in the Time Domain\label{sec:3-Testing}}
\subsection{Experimental Setup \label{sec:3-Experiment}}
\begin{figure}
\begin{center}
\includegraphics[width=3in]{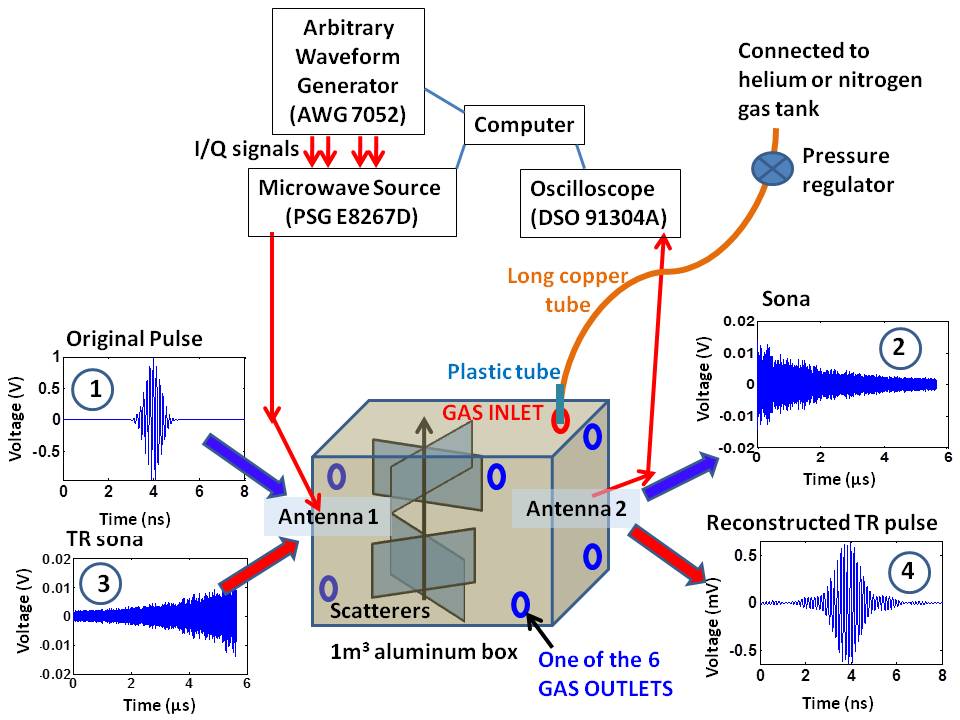}
\end{center}
\begin{quote}
\caption{Schematic of the experimental set up to induce VCSPPs in an electromagnetic cavity, and the equipment needed to implement an electromagnetic time reversal mirror. The VCSPP is induced by filling the cavity with helium or nitrogen gas. The gas transfer is carried out using a long copper tube that helps to warm the gases up to room temperature. There is a gas inlet, and six gas outlets on the walls of the cavity. The cavity has two antennas that are connected to a microwave source and an oscilloscope. The electromagnetic time reversal is carried out as follows. First, the original pulse is broadcast through antenna 1 (1), and the resulting sona is collected at antenna 2 (2). Next, the time reversed sona is injected into the system at antenna 1 (3) utilizing spatial reciprocity to retrieve the reconstructed time reversed pulse at antenna 2 (4). Experimental data are shown for each step. \label{fig:vcpFig2}}
\end{quote}
\end{figure}

The cavity that is used to test the sensing techniques is an approximately $1m^{3}$ (i.e. dimensions of $1.27m$ x $1.27m$ x $0.65m$) aluminum box that has scatterers and interior surface irregularities which facilitate the creation of ray chaotic trajectories (see Fig.~\ref{fig:vcpFig2}). The cavity is a mixed chaotic and regular billiard system because it has parallel walls which may support integrable modes in addition to the chaotic modes. Overall, the cavity represents a real world case in which the sensor would operate. There are two ports that connect the cavity to a microwave source and an oscilloscope. Each port consists of a monopole antenna of length $\approx1cm$, and diameter $\approx1mm$. The monopole antennas are mounted on two different walls of the cavity. An electromagnetic pulse with a center frequency of $7GHz$, and a Gaussian envelope of standard deviation $1ns$ is typically broadcast into the cavity through port 1. The resulting sona signal is collected at port 2 by the oscilloscope, and it is digitally filtered to minimize noise.

Experimentally inducing a VCSPP can be more challenging than inducing a VCP, which may slightly change the shape of the enclosure. VCSPPs can be realized by changing the speed of wave propagation within the cavity (i.e. changing the electrical volume). The speed of light (for the electromagnetic experiment at $7GHz$) within the cavity can be changed by filling up the cavity with different gases which have similar dissipation and dispersion properties. For example, the relative dielectric constant ($\epsilon_{r}$) of air (at $\approx50\%$ relative humidity), nitrogen gas, and helium gas are $\epsilon_{r, air}=1.000635$, $\epsilon_{r, N_{2}}=1.000547$, $\epsilon_{r, He}=1.000065$, respectively, at a temperature of $20^{0}C$ and a pressure of $101 kPa$ \cite{Essen1953, Newell1965}. As discussed later in Sec.~\ref{sec:3-FDTD}, it was shown that the slightly different dissipation values of these gases does not change the sona signal in any perceivable way. However, the slightly different speed of light values of these gases was seen to significantly change the sonas collected.

The experimental procedure for filling up the cavity with different gases is as follows (see Fig.~\ref{fig:vcpFig2}). There is a gas inlet on the top wall of the cavity. The gas inlet was connected to a gas tank via a plastic tube and a long copper tube coil. The long copper tube coil allowed the gas to reach room temperature before it gets into the cavity. There was a pressure regulator in between the gas tank and the copper tube to control the rate of flow of the gas. There were three gas outlets near the top wall of the cavity, and three gas outlets near the bottom wall of the cavity. The diameter of the gas inlet and outlets was about a fifth of the wavelength, so that no significant microwave leakage occurred. Depending on the density of the gas which was being pumped into the cavity, half of the outlets (near the top or bottom wall) were closed off with tape. This procedure helped to displace the existing gas and retain the gas being pumped into the cavity.

As detailed in Sec.~\ref{sec:3-Results}, sona signals were collected from the cavity both during and after the gas transfer process. The sona signals that were collected during the gas transfer indicate when the cavity is almost fully filled with the new gas. The sona signals that were collected after the gas transfer were used to measure the VCSPP that was induced.

\subsection{Finite Difference Time Domain (FDTD) Simulations \label{sec:3-FDTD}}
The experiment described in Sec.~\ref{sec:3-Experiment} was modeled using a Finite Difference Time Domain (FDTD) code. The FDTD code solves Maxwell's Equations inside a 3D numerical model of the experimental cavity introduced in Sec.~\ref{sec:3-Experiment}. The code for this simulation was optimized for parallel computers and for the simulation of reverberation chambers \cite{Moglie2004, Moglie2011}. The FDTD simulation of the cavity enabled a direct comparison of experimental and simulation results.

The FDTD simulation formed a 3D model of the experimental cavity by using spatial cubic cells with an edge length of $\Delta x\approx3.49mm$, and the cavity consisted of $364*364*188$ cells. The smallest time step taken to propagate solutions of Maxwell's Equations through the cells was $\Delta t=6ps$. Therefore, the Courant's number for computational stability was $\frac{c\Delta t\sqrt{3}}{\sqrt{\epsilon_{r}}\Delta x}\approx0.89<1$ for all the media ($\epsilon_{r}$) considered, where $c$ is the speed of light in vacuum \cite{Taflove2000}. The model of the cavity also had two ports, with antennas that are similar to the monopoles used in the experiment. The electromagnetic pulse broadcast into the model was also similar to the one broadcast experimentally (i.e. center frequency of $7GHz$, and width of $1ns$). The maximum and minimum of the ratio of the wavelength to the cubic cell dimension were $14$ and $11$ respectively.

In the experiment, the main source of dissipation is ohmic loss from the aluminum walls of the cavity. In the simulation, the walls were assumed to be lossless for simplicity. Instead, an equivalent loss was introduced within the medium of wave propagation to achieve the same quality factor as the experimental cavity \cite{Moglie2004}. To accomplish this, a uniform conductivity of $10^{-5}S/m$ was introduced throughout the interior of the cavity model.

As mentioned in Sec.~\ref{sec:3-Experiment}, the experiment relies on changing the electrical volume (at $7GHz$) within the cavity by changing the gases filling the cavity. The gases used were air, nitrogen gas, and helium gas. For typical laboratory atmospheric conditions, the electromagnetic loss in air at $7GHz$ can be mainly attributed to oxygen molecules \cite{Rohan1991, Moglie2004}. The specific attenuation of oxygen is $0.007dB/km$ whereas the specific attenuation of water vapor is just $0.003dB/km$ at $7GHz$. The conductivity of air is estimated to be $4.3*10^{-9}S/m$ \cite{Moglie2004}. Therefore, the dissipation inside the cavity filled by helium gas or nitrogen gas was roughly modeled by introducing an equivalent loss of $10^{-5}S/m - 4.3*10^{-9}S/m = 9.9957*10^{-6}S/m$ uniformly throughout the cavity (once again, the walls were assumed to be lossless in the model). It was shown that sonas that are collected from the cavity model with $10^{-5}S/m$ conductivity, and sonas that are collected from the cavity model with $9.9957*10^{-6}S/m$ are identical (i.e. they have a fidelity of $1$). This simulation result proved that the difference in loss among air (at $\approx50\%$ relative humidity), nitrogen gas, and helium gas is not significant.

As will be discussed later in Sec.~\ref{sec:3-Results}, the gases that experimentally filled the cavity are not pure. The effective $\epsilon_{r}$ of the gases that filled the cavity are different from the $\epsilon_{r}$ values for pure gases. The simulation of the cavity that is filled with helium, nitrogen or air is done by using the effective $\epsilon_{r}$ values. This allows the simulation to better model the experimental reality. The $\epsilon_{r}$ values used to simulate a cavity filled with air, nitrogen and helium are $1.000576$, $1.000547$, and $1.00017$ respectively. These values were chosen to match the experimental results that will be discussed in Sec.~\ref{sec:3-Results}. The values were determined by anchoring the effective $\epsilon_{r}$ of nitrogen to the value for pure nitrogen gas, and finding the effective $\epsilon_{r}$ values for the other gases to match the experiment.

\subsection{Sensitivity of the Quantitative Sensor \label{sec:3-Sensitivity}}
Here, the sensitivity of the sensor is described by deriving an expression for the minimum volume changing perturbation that can be measured. Then, it will be shown that the VCSPP which is induced when either of the three gases in the experiment (i.e. air, nitrogen and helium gas) is displaced by another one, can be measured using our experimental set up.

Consider using the sensing technique based on scattering fidelity, which was introduced in Section~\ref{sec:3-Theory}. The technique relies on comparing the baseline and perturbed sonas generated by a short pulse that excites many modes of the system as a function of time. In Sec.~\ref{sec:3-Sensitivity}, it is assumed that the perturbation increases the volume by a factor of $P^{3}$, where $P>1$. At time $t=0$, the sonas are expected to be similar for small enough $\Delta t$ (see Eq.~\ref{eqn:sf}), hence $SF(t=0)\approx1$. At any other time $t$, there may be a perceptible difference between the sonas. Any particular signal feature in the baseline sona at time $t$ is expected to be seen in the perturbed sona at time $t+t_{gap}(t)$, where $t_{gap}(t)=P \; t-t$; here, $t_{gap}(t)$ is defined as the time gap that develops between two identical features in the baseline and perturbed sona at time $t$, where $t$ is measured within the baseline sona. This is because the baseline sona stretched out along its time axis by a factor of $P$ should approximate the perturbed sona, as discussed in Sec.~\ref{sec:3-Theory}.

The minimum volume changing perturbation that can be quantified is the minimum of $P^{3}=(\frac{t+t_{gap}(t)}{t})^{3}$. On the other hand, when the $SF$ of the baseline and perturbed sona is computed at any time $t$ (see Eq.~\ref{eqn:sf}), there are two necessary conditions that should be satisfied in order to be able to measure the perturbation. First, the signal-to-noise-ratio ($SNR$) of the baseline sona at time $t$ should be well above $1$. This is because the $SF$ of the baseline and perturbed sonas (when the $SNR$ is close to 1) would simply be the correlation of two noisy signals. Second, $t_{gap}(t)$ should be, conservatively, greater than half of the period of the oscillations in the sona signals. Otherwise, if $t_{gap}(t)$ is much smaller than half a period of the sona oscillations, the $SF(t)$ will not be convincingly lower than its maximum value at $t=0$ (which is $\approx1$ for the appropriate $\Delta t$ value in Eq.~\ref{eqn:sf}), hindering reliable measurement of the perturbation. These two conditions guarantee that the $SF(t)$ of the baseline and perturbed sona can be used to measure the perturbation. The minimum of $P^{3}=(\frac{t+t_{gap}(t)}{t})^{3}$ subject to these two conditions is $(1+\frac{T/2}{T_{NoiseLevel}})^{3}$, where $T$ is the period of sona oscillations and $T_{NoiseLevel}$ is the time at which the $SNR$ of the sona approaches $1$.

Therefore, the minimum perturbation that can be quantified by the sensor (i.e. $(1+\frac{T/2}{T_{NoiseLevel}})^{3}$) depends on the wavelength of the waves used to probe the cavity, the dissipation in the cavity and the $SNR$ of the system. The $SNR$ in turn depends on the noise in the system, and the dynamic range of the wave generation and detection equipment. The equipment used in this experiment are shown in Fig.~\ref{fig:vcpFig2}. For the electromagnetic experimental set up that is probing the $1m^{3}$
cavity discussed in this work, using $5cm$ wavelength waves, a VCSPP that is as small as $4$ parts in $10^{5}$ can be measured. Note that it is assumed that the probing pulse excites several resonances of the baseline system as it is discussed further in Sec.~\ref{sec:3-LimitationTD}.

Getting back to the experimental set up shown in Fig.~\ref{fig:vcpFig2}, the VCSPP was induced by changing the gas filling the cavity from air to nitrogen gas or to helium gas. This results in $\approx0.05\%$ change in the $\epsilon_{r}$ of the gases, which is equivalent to a VCSPP of at least $13$ parts in $10^{5}$. Therefore, the experimental system is expected to detect the change in electrical volume induced when one of these gases displaces the other inside the cavity (assuming that the gases are pure).

\subsection{Results from the Experiment and the FDTD Simulation \label{sec:3-Results}}
\subsubsection{Sensing Using Scattering Fidelity \label{sec:3-ResultsSF}}
\begin{figure}
\begin{center}
\includegraphics[width=3in]{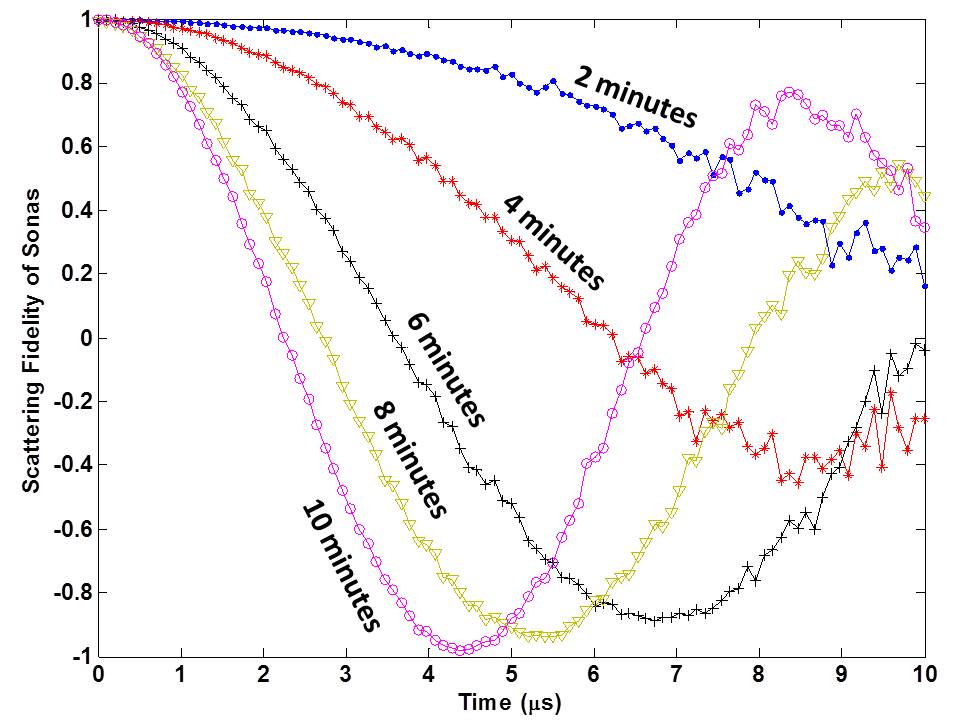}
\end{center}
\begin{quote}
\caption{Scattering fidelity of sona signal from a cavity that is filled with air (baseline), and sona from a perturbed cavity that had $2$, $4$, $6$, $8$, and $10$ minutes of nitrogen gas inflow. The perturbation gets stronger as the concentration of nitrogen gas increases in the perturbed cavity. \label{fig:vcpFig3}}
\end{quote}
\end{figure}

The air in the cavity at $\approx20^{0}C$ and $\approx50\%$ relative humidity, was systematically displaced with nitrogen gas at room temperature. The nitrogen gas was pumped into the cavity at $207kPa$ gauge pressure as the air flowed out through the gas outlets of the cavity. Every two minutes, the flow was stopped, and $10$ nominally identical sonas (which are actually almost identical) were measured from the cavity, and these were averaged together. The averaging is done after aligning the sonas to eliminate the adverse effects of trigger jitter in the data acquisition system. In this manner, five averaged sonas were collected from the cavity as the cavity was filled with more and more pure nitrogen gas. Each of these five sonas were compared with a sona that was collected from the original cavity filled with air. The comparison was done by computing the scattering fidelity (see Eq.~\ref{eqn:sf}) of the sona from airy cavity and a sona from a partially air-filled cavity. Fig.~\ref{fig:vcpFig3} shows these scattering fidelities. The concentration of nitrogen increases with the number of minutes of nitrogen inflow into the cavity. Therefore, Fig.~\ref{fig:vcpFig3} shows scattering fidelities of VCSPPs which get progressively stronger.

\begin{figure}
\begin{center}
\includegraphics[width=3in]{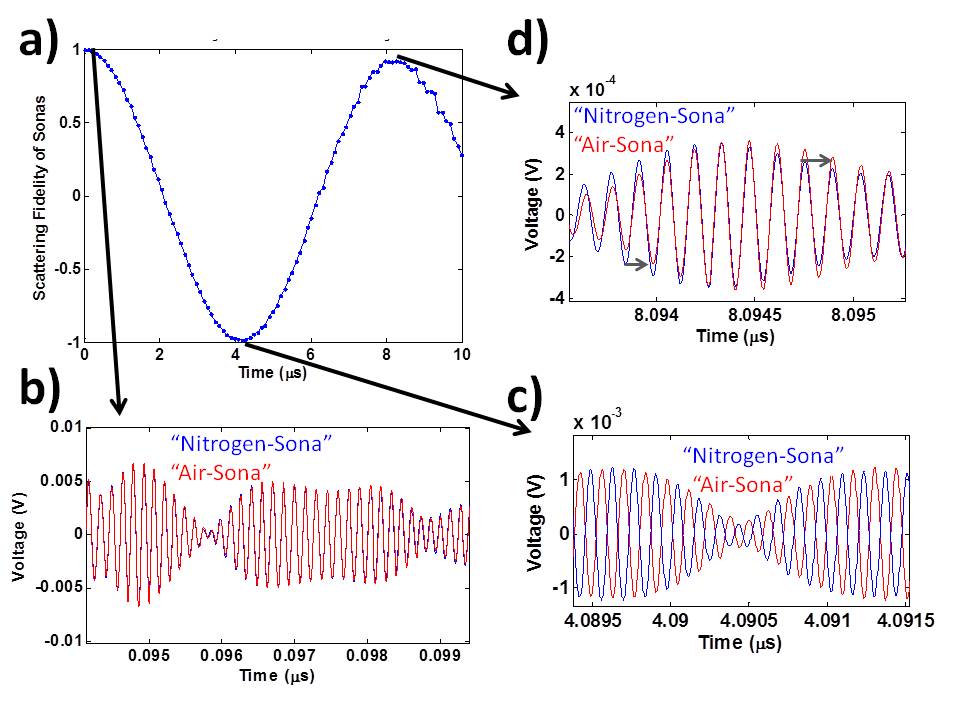}
\end{center}
\begin{quote}
\caption{(a) Examining the scattering fidelity oscillation for VCSPPs using sona from air filled cavity and sona from nitrogen filled cavity. Each of the sonas are averaged over $100$ sona samples. (b) The sonas near $t=0$ have fidelity of $1$. (c) The sonas are out of phase by half a period around $t=4\mu s$. (d) The sonas are out of phase by a period around $t=8\mu s$. Besides, the phase shift between the envelopes of the sonas becomes significant. \label{fig:vcpFig4}}
\end{quote}
\end{figure}

The scattering fidelity of a VCSPP shows oscillation whose period is inversely related to the strength of the perturbation (see Fig.~\ref{fig:vcpFig3}). The oscillation in the scattering fidelity can be explained by specifically examining the scattering fidelity of sona from the air filled cavity, and sona from the nitrogen gas filled cavity (see Fig.~\ref{fig:vcpFig4}(a)). Here, both of the sonas were obtained by averaging over $100$ nominally identical sona samples. For times relatively close to $t=0$, the difference in the speed of light between air and nitrogen does not show up when signals that traveled through these two gases are compared (see Fig.~\ref{fig:vcpFig4}(b)). However, at $t\approx 4\mu s$, a phase shift of half a period (i.e. $0.07ns$ for the probing pulse centered at $7GHz$) develops between the two signals (see Fig.~\ref{fig:vcpFig4}(c)). Thus, the scattering fidelity between the two sonas becomes $\approx-1$. The relative phase shift between the two sonas increases to one full period at about $8\mu s$, and hence the scattering fidelity recovers to almost $1$ (see Fig.~\ref{fig:vcpFig4}(d)). However, as can be seen in Fig.~\ref{fig:vcpFig4}(d), in addition to the phase shift that develops between the individual oscillations of the two sonas, a relative phase shift starts to develop between the envelopes of the sona signals. Thus, the scattering fidelity of VCSPPs is not expected to oscillate between $1$ and $-1$ indefinitely. Rather, it is expected to decay while oscillating \cite{Gorin2006a}. However, we do not anticipate seeing this fidelity decay with our measurement system for this particular perturbation because the $SNR$ of the sonas approaches unity after about $10\mu s$, and the fidelity does not decay appreciably within this time for this particular perturbation.

\begin{figure}
\begin{center}
\includegraphics[width=3in]{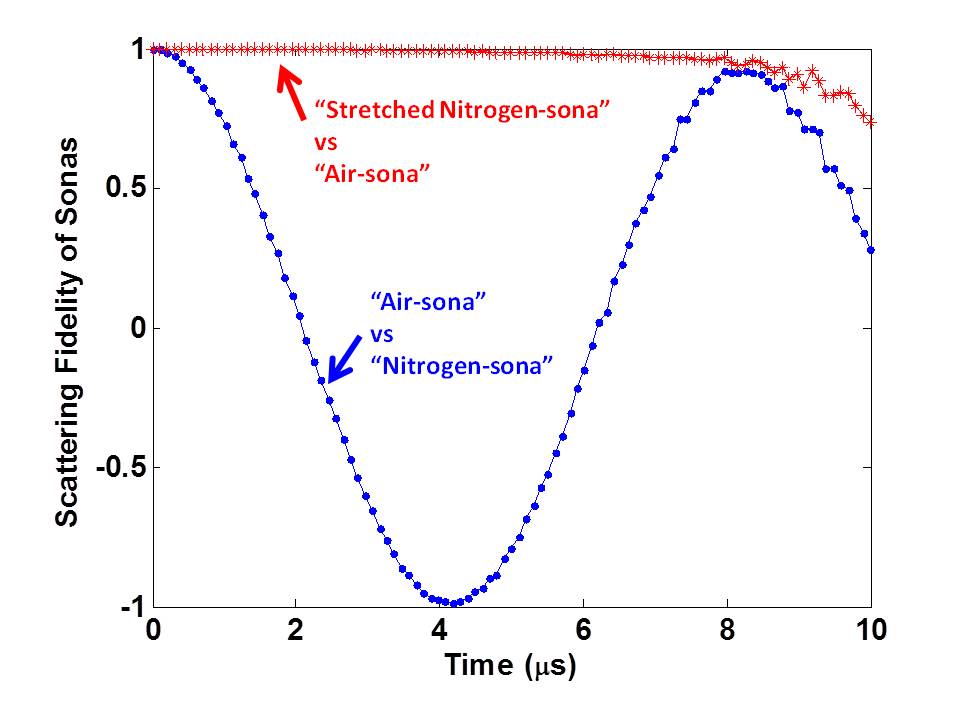}
\end{center}
\begin{quote}
\caption{Undoing the effect of a VCSPP. The sona that was collected from the cavity filled with nitrogen was stretched out optimally to recover the scattering fidelity to $1$ throughout the times when the $SNR$ of the sonas is robust. The SNR decreases roughly by a factor of a thousand between $0\mu s$ and $10\mu s$. The optimum stretching factor quantified the VCSPP. The fact that the scattering fidelity was recovered proved that the perturbation was VCSPP. \label{fig:vcpFig5}}
\end{quote}
\end{figure}

The speed of light at $7GHz$ in pure nitrogen gas was faster than it was in the laboratory air (which had about $50\%$ relative humidity). Therefore, as discussed in Sec.~\ref{sec:3-Theory}, the VCSPP can be quantified by finding the optimum stretching factor to be applied on the sona that is collected from the cavity filled with nitrogen. The goal is to recover the scattering fidelity of the stretched "nitrogen-sona" and the "air-sona" to $1$. This was achieved by using a stretching factor given by $(4\mu s+0.07ns)/4\mu s$; because, $t_{gap}=0.07ns$ ($\approx$ half a period of oscillation of the sona) at $t=4\mu s$ ($SF\approx-1$) based on the discussion in Sec.~\ref{sec:3-Sensitivity}. The resulting scattering fidelity is plotted in red (see Fig.~\ref{fig:vcpFig5} ) with the scattering fidelity of the unmodified sonas which is shown in blue. The optimum stretching factor is $(4\mu s+0.07ns)/4\mu s\approx1.000018$. Whereas, $\sqrt{\frac{\epsilon_{r,Air}}{\epsilon_{r,N_{2}}}}\approx1.000044$ is the expected value of $P$. In other words, ideally a $44ppm$ (parts per million) change across each electrical dimensions of the cavity is expected, however, an $18ppm$ change is measured.

The discrepancy can be explained by the fact that the nitrogen gas that filled the cavity has an effective $\epsilon_{r}$ that is probably much closer to the effective $\epsilon_{r}$ of air. Fig.~\ref{fig:vcpFig3} shows that it can take several minutes to displace the air in the cavity with pure nitrogen. Therefore, we do not expect the $\epsilon_{r}$ of the nitrogen gas that filled the cavity to be the same as the literature value for pure nitrogen gas. The same conclusion applies to the other gases filling the cavity. There are several reasons for the expected discrepancy between the literature value of $\epsilon_{r}$ and the effective $\epsilon_{r}$ of the gases filling the cavity experimentally. The helium and nitrogen gases are of industrial quality, which is not perfectly pure. Besides, the cavity is not necessarily air tight; this could lead to leakage of air into the cavity when the cavity is not pressurized. Finally, the $\epsilon_{r}$ of the gases is a function of temperature and relative humidity (for the case of air) \cite{Newell1965}. Therefore, deviations of laboratory temperature and relative humidity from $20^{0}C$ and $50\%$ could impact the effective $\epsilon_{r}$ value of the gases.

The fact that the shape of the cavity was preserved (during the displacement of the air by nitrogen gas) can also be seen from Fig.~\ref{fig:vcpFig5}. If the shape of the cavity were not preserved, it would not be possible to recover the scattering fidelity of the two sonas through simple numerical stretching of one of the sonas, along the time axis. Hence, displacing air with nitrogen gas is not just a VCP, but is also a VCSPP. Later, a VCP induced by displacement of air with helium, which is not VCSPP, will be discussed.

\begin{figure}
\begin{center}
\includegraphics[width=3in]{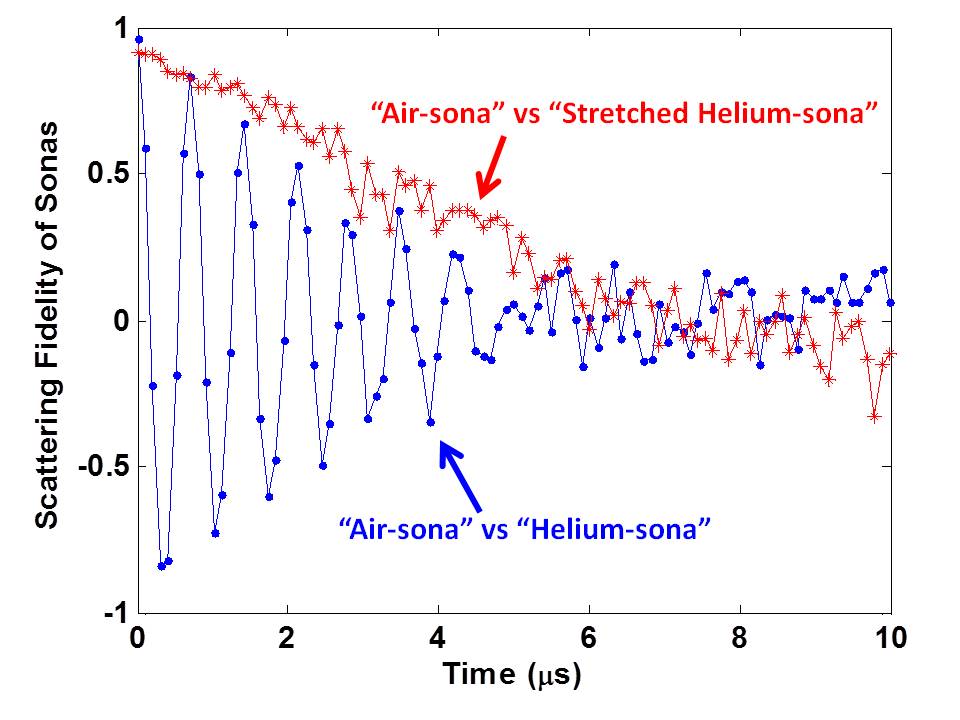}
\end{center}
\begin{quote}
\caption{Scattering fidelity of sona from a cavity filled with air and sona from a cavity filled with helium. The fidelity oscillations and decay can be seen. The effect of the VCP could not be completely undone by stretching the "helium sona", so, the perturbation is not VCSPP. The buoyant force of helium gas can slightly change the shape by flexing the walls of the cavity. \label{fig:vcpFig6}}
\end{quote}
\end{figure}

The fidelity decay of VCSPPs, which is expected to be superimposed on the fidelity oscillations, can be seen for a stronger VCSPP.  The VCSPP should be strong enough to bring about a significant phase shift between the envelopes of the sonas before their $SNR$ deteriorates. For our experimental set up, such a strong VCSPP can be achieved by displacing the air in the cavity by helium gas. The scattering fidelity of sona from a cavity that is filled with air and sona from a cavity that is filled with helium is plotted in Fig.~\ref{fig:vcpFig6} in blue. Based on the definition in Sec.~\ref{sec:3-Sensitivity}, $t_{gap}=0.07ns$ (which is half the period of sona oscillation) at time $t=0.35\mu s$. Thus, the optimum stretching factor was chosen to be $(0.35\mu s+0.07ns)/0.35\mu s=1.0002$. This optimum value maximizes the average value of the resulting scattering fidelity. Since the speed of light at $7GHz$ is higher in helium than in air, it was the "helium sona" that was stretched out along its time axis. The scattering fidelity of the stretched "helium sona" and the "air sona" is also shown in Fig.~\ref{fig:vcpFig6} in red. Once again, the stretching factor $(0.35\mu s+0.07ns)/0.35\mu s=1.0002$ approximates $\sqrt{\frac{\epsilon_{r,Air}}{\epsilon_{r,He}}}\approx1.000285$. In other words, a $285ppm$ change across each electrical dimensions of the cavity is expected, and a $200ppm$ change is measured. This shows that the change in electrical volume which was induced by replacing the air in the cavity with helium was quantified successfully, considering the fact that the effective $\epsilon_{r}$ of the helium and air gases are probably closer than expected.

However, unlike the case in Fig.~\ref{fig:vcpFig5}, the effect of the VCP could not be undone perfectly. The scattering fidelity of the stretched "helium sona" and the "air sona" was not close to $1$ throughout time; instead it shows a fidelity decay. The scattering fidelity of "helium sona" and "air sona" is expected to oscillate between $1$ and $-1$ (as the phase shift between the fast oscillations of the sonas increases in a similar fashion to the illustration in Fig.~\ref{fig:vcpFig4}(b-d)) and decay to $0$ (as the phase shift between the complicated envelopes of the sonas increases). This decay is seen experimentally in Fig.~\ref{fig:vcpFig6}. However, there must be another fidelity decay superimposed on the fidelity decay that can be attributed to a VCP; because the fidelity decay could not be undone by numerically stretching out one of the sonas.

\begin{figure}
\begin{center}
\includegraphics[width=3in]{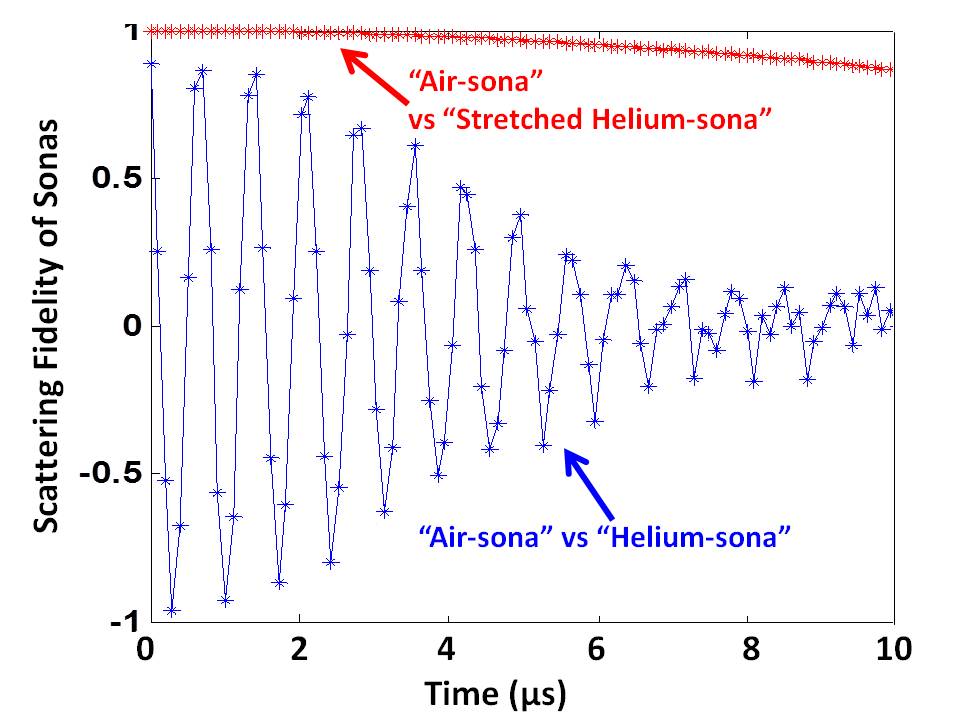}
\end{center}
\begin{quote}
\caption{Scattering fidelity of sona from a cavity filled with air and sona from a cavity filled with helium. The sonas are generated from the FDTD model of the cavity. \label{fig:vcpFig7}}
\end{quote}
\end{figure}

A FDTD simulation of the experiment was performed to better understand the form of the scattering fidelity when comparing "helium sona" and "air sona". As discussed in Sec.~\ref{sec:3-FDTD}, differences in the dissipation of helium gas and air are so minute that they do not need to be considered. The scattering fidelity of "helium sona" and "air sona", which were obtained from the FDTD model by broadcasting the $7GHz$ pulse used experimentally, are shown in Fig.\ref{fig:vcpFig7}. The simulation results show that the effect of the VCSPP can be undone by applying the optimum stretching factor (i.e. $(0.35\mu s+0.07ns)/0.35\mu s=1.0002$) to the "helium sona". The scattering fidelity of the "air sona" and the stretched "helium sona" is shown in red in Fig.\ref{fig:vcpFig7}; it is close to $1$ which shows that the effect of the perturbation can be undone by simple numerical stretching. The fidelity is close to $1$ for times where the numerical errors in the FDTD are negligible.

The difference in the results in Fig.~\ref{fig:vcpFig6} and Fig.~\ref{fig:vcpFig7}, shows that experimentally displacing air with helium induces another kind of perturbation other than a volume changing perturbation. The additional perturbation was not seen when nitrogen was pumped into the cavity at the same pressure setting (i.e. $207kPa$ gauge pressure) as was used for pumping helium into the cavity. One possible explanation for this discrepancy is the fact that helium gas exerts a significant buoyancy force which slightly flexes the walls of the cavity (about $1m^{2}$ area, and $3mm$ thick aluminum sheets). From previous work on this cavity, it was shown that a flexing of one of the walls of the cavity can cause a significant shape changing perturbation \cite{Anlage2007}. The result shown in Fig.~\ref{fig:vcpFig6} demonstrates that it is possible to verify if the shape of the cavity remained intact while its electrical volume changed. This verification can be simply done by checking if the scattering fidelity can be recovered to $1$ throughout time. The capability to detect changes to the shape of the cavity during a volume changing perturbation (which could be induced by a spatially uniform heating or cooling of a homogenous cavity) can have several applications as was pointed out in Sec.~\ref{sec:3-Introduction}. Yet another possible explanation for the above mentioned discrepancy is the fact that nitrogen and helium gases have different density and atomic sizes. This can lead to differences in the spatial uniformity of the gasses filling the cavity. Again, such differences can have the character of cavity shape changing perturbations.

\begin{figure}
\begin{center}
\includegraphics[width=3in]{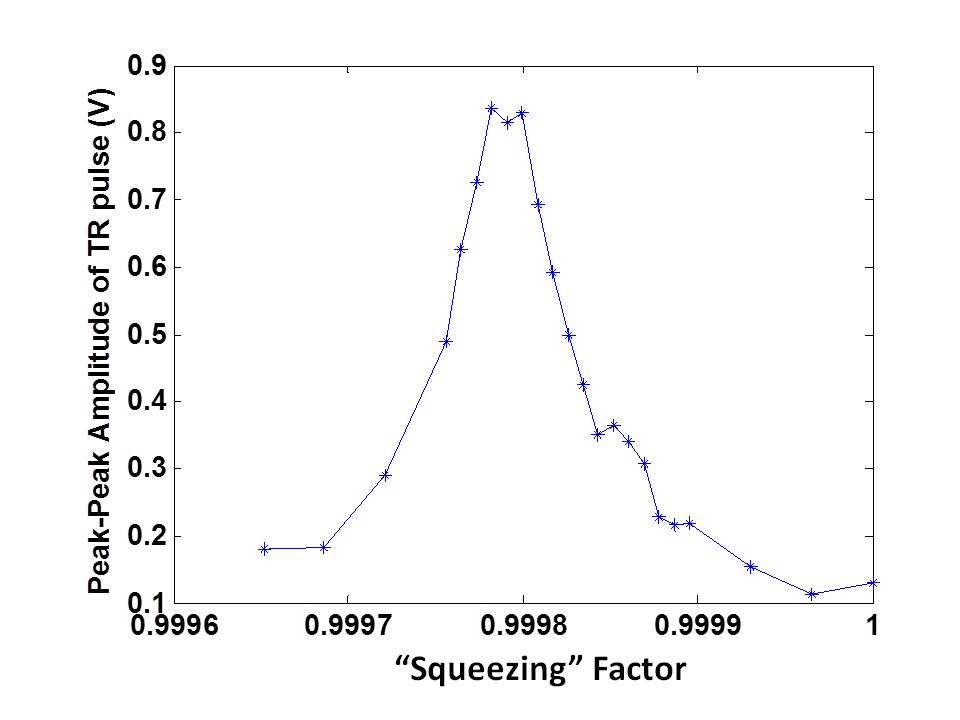}
\end{center}
\begin{quote}
\caption{Plot of peak-to-peak amplitude of time-reversed pulse in a cavity filled with Helium vs "squeezing" factor used to scale the time axis of the sona which is collected from a cavity filled with Nitrogen. \label{fig:vcpFig8}}
\end{quote}
\end{figure}

\subsubsection{Sensing Using Time Reversal \label{sec:3-ResultsTR}}
So far, only one of the sensing techniques introduced in Sec.~\ref{sec:3-Theory} is demonstrated to quantify volume changing perturbations. In addition to the scattering fidelity technique, time reversal mirrors can be used to quantify volume changing perturbations. When a sona signal that is collected from the cavity filled with nitrogen is time reversed and broadcast into the cavity filled with helium, the reconstructed time reversed pulse is not expected to be ideal. This is because the time reversed sona would be traversing an effectively smaller cavity. Note that in this particular case, the baseline cavity is electrically larger than the perturbed cavity. Based on the discussions in Sec.~\ref{sec:3-Theory}, the baseline sona should be squeezed along its time axis using an optimum factor to recover the maximum amplitude time reversed pulse. When the optimally squeezed "nitrogen sona" is time reversed and broadcast into the cavity filled with helium, the reconstructed time reversed pulse is expected to better approximate the original pulse.

The improvement in the quality of the time reversed reconstructed pulse can be measured in various ways \cite{Taddese2010}. Here, the simplest measure (the peak-to-peak-amplitude (PPA) of the reconstructed pulse) of the quality of the time reversed pulse is used. Fig.~\ref{fig:vcpFig8} shows the PPA of the time reversed reconstructed pulse obtained when the "nitrogen sona" is numerically squeezed along its time axis by varying amounts. The optimum squeezing factor of $0.9998$ approximates $\sqrt{\frac{\epsilon_{r, He}}{\epsilon_{r, N_{2}}}}\approx0.99976$. This means that a $240ppm$ change is expected, but a $200ppm$ change across each electrical dimensions of the cavity was measured. Once again, the small discrepancy can be explained by the fact that the gases in the cavity are not perfectly pure. This shows that the time reversal technique can also be used to quantify volume changing perturbations.

Time reversal mirrors can also be used to detect when a volume changing perturbation slightly changes the shape of the cavity. In this case, when the "nitrogen sona" is time reversed and broadcast into the cavity filled with nitrogen, the PPA of the reconstructed time reversed pulse was about $1.13V$. However, when an optimally squeezed "nitrogen sona" is broadcast into the cavity filled with helium gas, the optimal value of the PPA is only about $0.85V$. This indicates that the perturbation, which is induced when the nitrogen gas inside the cavity is displaced with helium gas, is not just a volume changing perturbation but also a shape changing perturbation. Once again, this is due to the relatively strong buoyant force of helium gas which can flex the walls of the cavity.

To summarize, either of the two time domain sensing techniques introduced in Sec.~\ref{sec:3-Theory} can be used to identify and quantify a VCSPP. However, the sensing technique based on time reversal can have an advantage because it is computationally cheaper \cite{Taddese2010}.

\subsection{Limitation of the Time Domain Approaches to Measure VCSPP \label{sec:3-LimitationTD}}

Now we can describe the time domain approach further, and point out its limitations. When a cavity is monitored for a VCSPP, a pulse is periodically broadcast into it, and a sona is collected. It is generally preferred to use the same probing pulse (i.e. the same center frequency, bandwidth, amplitude, and shape) so that only the changes in the system show up when looking at the sonas. When a VCSPP (with $P>1$) occurs, the transmission spectrum $|S_{12}|^{2}$ of the system shrinks along the frequency axis by a factor of $P$ as shown in Figs.~\ref{fig:vcpFig1}(c)\&(d). The probing pulse frequency coverage is schematically shown in the frequency domain in Figs.~\ref{fig:vcpFig1}(c)\&(d). The baseline and the perturbed sonas are a result of the probing pulse exciting resonances of the cavity. The resonances excited by the probing pulse in the baseline and perturbed cavity are not all the same. Suppose that there is a significant overlap between the resonances excited by the probing pulse in the baseline and perturbed cavity. Under this condition, we expect that the baseline sona can be numerically stretched out by a factor of $P$ along its time axis to approximate the perturbed sona.

However, if there is no overlap between the resonances excited in the baseline and perturbed system, the time domain sensing techniques are not expected to work. Thus, both of the time domain sensing techniques (based on scattering fidelity and time reversal mirrors) face a limitation regarding the maximum VCSPP that can be measured. This limitation can be improved by increasing the bandwidth of the pulse that is used to probe the cavity. Doing so would effectively make the time domain technique closer to the frequency domain interrogation of the system, which does not face a limit on the maximum VCSPP that can be measured. In a similar spirit, using a pulse with a flat frequency response (such as a chirp) may also be helpful. Note that the frequency domain interrogation of the system is assumed to be done on a large enough frequency window to measure the perturbation.

The regime of VCSPP strengths that cannot be measured using the time domain techniques is estimated as follows. Suppose that the probing pulse excites resonances with frequencies ranging from $f_{min}$ to $f_{max}$, with the center frequency at $f=\frac{f_{min}+f_{max}}{2}$, and $3dB$ bandwidth of $f_{max}-f_{min}$. It is assumed that the probing pulse excites several resonances of the baseline system. In other words, $f_{max}-f_{min} \gg \Delta f$, where $\Delta f$ is the mean spacing between the resonant frequencies of the baseline cavity. The VCSPP changes the volume of the cavity by a factor of $P^{3}$, where $P>1$ for simplicity. Once again, the VCSPP has the effect of scaling the $|S_{12}|^{2}(f)$ of the cavity along the frequency axis by a factor of $P$.  If $\frac{f_{max}}{P} < f_{min}$, then the VCSPP is not expected to be measured by using this probing pulse in the time domain because there would be significant overlap between the resonances excited in the baseline and perturbed cavities.

In the experiments discussed in Sec.~\ref{sec:3-Experiment}, the probing pulse excited resonances between $f_{min}=6.5GHz$ and $f_{max}=7.5GHz$. The mean spacing between the resonant frequencies of the baseline cavity is given by $\Delta f = \frac{c^{3}}{8\pi V f^{2}} = \frac{\lambda^{3}}{V}\frac{f}{8\pi}\approx22kHz$; where $V$ is the volume of the baseline cavity, $f=7GHz$ is the center frequency of the pulse, and $\lambda$ is the wavelength. Therefore, a VCSPP where $P > \frac{f_{max}}{f_{min}} \approx 1.15$ is not expected to be measured in the time domain experiments discussed in Sec.~\ref{sec:3-Experiment}. Clearly, the strongest VCSPP that was achieved in the laboratory (i.e. $P\approx1.000018$ for nitrogen gas vs the air), and the strongest VCP (i.e. $P\approx1.0002$ for helium gas vs the air) are both far below the maximum perturbation strengths that can be measured using the time domain techniques.

The case of a strong VCSPP perturbation, which cannot be measured using the time domain techniques, is best considered using a simulation tool that can be easily interrogated in the frequency domain as well. Sec.~\ref{sec:3-FreqDomain} discusses such strong perturbations, and shows how they can be quantified using a frequency domain approach.

\section{Sensing Using Frequency Domain Information \label{sec:3-FreqDomain}}
In Sec.~\ref{sec:3-Theory}, it was mentioned that VCSPPs can be quantified using information obtained in the time domain or frequency domain. The time domain approach is generally more practical in applications. However, the frequency domain approach does not have limitations on the maximum perturbation value that can be quantified. This frequency domain approach is used to measure VCSPPs on a quasi-1D system called the star graph. We use the star graph because it is a type of quantum graph that has generic properties of wave chaotic systems \cite{Kottos2003}, but is relatively simple to understand. It is also computationally cheaper to simulate than the 3D wave chaotic system discussed in Sec.~\ref{sec:3-FDTD}. Besides, the star graph can be directly implemented in the frequency domain as discussed in Sec.~\ref{sec:3-StarGraph}.

\subsection{The Star Graph Model \label{sec:3-StarGraph}}
\begin{figure}
\begin{center}
\includegraphics[width=3in]{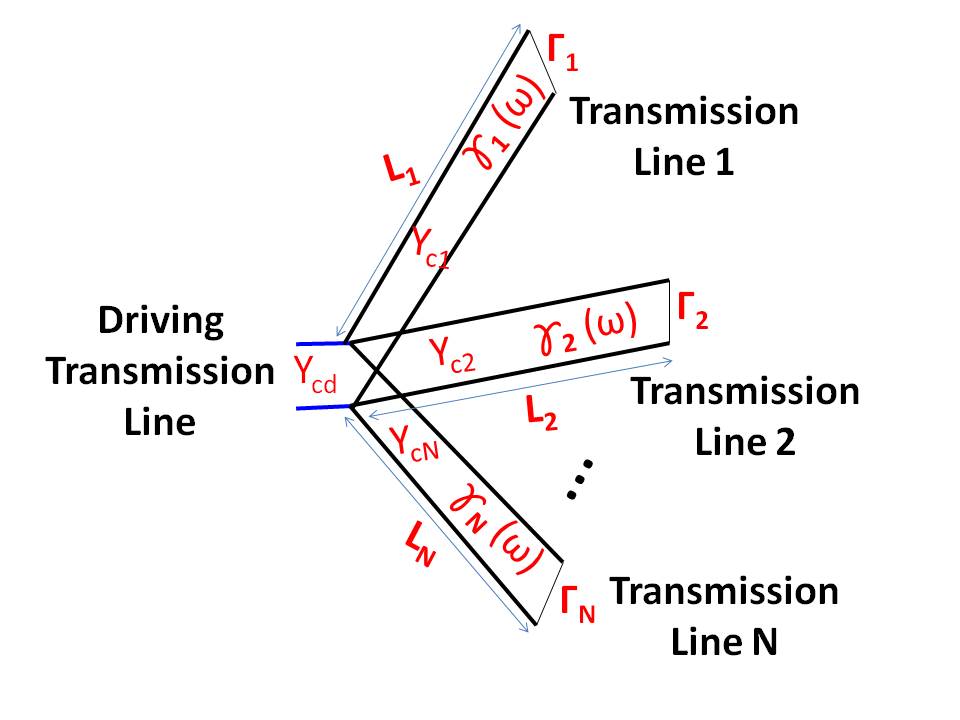}
\end{center}
\begin{quote}
\caption{Schematic of the star graph model. There are $N$ transmission lines that are connected in parallel, and a driving transmission line of zero length. Each of the $N$ lines (labeled by $n$) can have unique length ($L_{n}$), characteristic admittance ($Y_{cn}$), frequency dependent propagation constant ($\gamma_{n}(\omega)$), and reflection coefficient ($\Gamma_{n}$). The driving line has a characteristic admittance of $Y_{cd}$.  \label{fig:vcpFig9}}
\end{quote}
\end{figure}

The star graph is numerically modeled as a set of interconnected transmission lines as shown, schematically, in Fig.~\ref{fig:vcpFig9} \cite{Taddese2011}. This is a one port system, hence, the input signal is injected into the driving transmission line and the output signal is also retrieved from the same line. The driving transmission line has zero length. The driving transmission line is connected with a number of transmission lines which are all connected in parallel with each other. The transmission line properties (for a line labeled by $n$) are length ($L_{n}$), characteristic admittance ($Y_{cn}$), frequency dependent complex propagation constant ($\gamma_{n}(\omega)$), and complex reflection coefficient ($\Gamma_{n}$) (for reflection from the terminations of the lines that are not connected to the driving line). The driving line has zero length, thus its only adjustable property is its characteristic admittance ($Y_{cd}$).

This one port system is modeled by using the analytically derived expression for its scattering parameter as a function of frequency, $S_{11}(\omega)$. The scattering parameter can be expressed in terms of the characteristic admittance of the driving line ($Y_{cd}$) and the input admittance ($Y_{n}$) of each of the other transmission lines while looking towards them,
\begin{equation}
S_{11}(\omega) = \frac{Y_{cd} - \sum_{n=1}^{n=N}{Y_{n}(\omega)}}{Y_{cd} + \sum_{n}{Y_{n}(\omega)}},
\label{eqn:SofStarGraph}
\end{equation}
where $N$ is the number of transmission lines connected in parallel. The input admittance of each transmission line (labeled by $n$), $Y_{n}(\omega)$, can be expressed in terms of the above mentioned properties of the line,
\begin{equation}
Y_{n}(\omega) = Y_{cn}(\frac{1-\Gamma_{n}e^{2\gamma_{n}(\omega)L_{n}}}{1+\Gamma_{n}e^{2\gamma_{n}(\omega)L_{n}}}).
\label{eqn:InAdmiStarGraph}
\end{equation}

Once the scattering parameter of the system is computed over a broad frequency range, the response to any time domain input signal can be calculated. This is done by Fourier transforming the input signal to the frequency domain, multiplying it by the scattering parameter ($S_{11}(\omega)$) and inverse Fourier transforming the product back to the time domain. This establishes the star graph model as a time domain simulation of this quasi-1D system. However, in this section we are mainly interested in the frequency domain representation of the star graph using its scattering parameter. The frequency domain approach will be shown to be effective in quantifying strong perturbations that could not have been measured otherwise.

\subsection{Quantifying Strong Perturbations to the Star Graph \label{sec:3-QuantVolChaStarGraph}}
As discussed in Sec.~\ref{sec:3-Theory}, the scattering parameter of a cavity can be used to quantify a VCSPP. The star graph is a quasi-1D system. A perturbation that changes the length of all of the lines of the star graph by the same proportion (i.e. with constant $P$) changes its effective volume while leaving its shape intact (i.e. it is a VCSPP). The effective volume of a star graph with $N$ lines is given by $2\sum_{n=1}^{n=N}L_{n}$ \cite{Kottos1999}. Therefore, for a VCSPP perturbation that scales the lengths by a factor of $P$, the effective volume of the star graph also changes by a factor of $P$. This is different from the case of the 3D cavity discussed in Sec.~\ref{sec:3-Testing} (i.e. the volume of the 3D cavity changes by $P^{3}$).

The baseline star graph was set up using the following parameter values. There were $10$ lines whose length is given by $L_{n}=0.3\lceil10\sqrt{n}\rceil$ $m$ (for $n$ ranging from $1$ to $10$). Each of these lines had a characteristic admittance, $Y_{cn}$, of $1S$. The characteristic admittance of the driving line was chosen such that $Y_{cd}=\sum_{n=1}^{n=10}Y_{cn}=10S$ in order to eliminate prompt reflection of signals injected through the driving line. The propagation constant of the lines is a function of the frequency, $\omega$, and was given by $\gamma_{n}(\omega)=\imath\frac{\omega}{c}$ with $n=1,...,N$, where $c$ is the speed of light in vacuum; thus, the lines themselves were considered to be lossless (i.e. $\gamma_{n}(\omega)$ does not have a real part). However, energy was dissipated during reflection from the terminations of the lines. The reflection coefficient was given by $\Gamma_{n}=e^{\frac{-0.3\lceil10\sqrt{n}\rceil2}{c\tau}}$ with $n=1,...,N$, where $\tau=1.5\mu s$. The amount of dissipation was designed to be independent of the size of the star graph. However, the dissipation introduced through sub-unitary values of $\Gamma_{n}$ can be interpreted as an equivalent loss that could be introduced through $\gamma_{n}(\omega)$ (i.e. by introducing $\frac{1}{c\tau}$ as the real part of $\gamma_{n}(\omega)=\imath\frac{\omega}{c}+\frac{1}{c\tau}$). Thus,
if the dissipation were modeled using $\gamma_{n}(\omega)$, $\tau=1.5\mu s$ would be interpreted as the time it takes the signals to decay by $1/e$ as they propagate along the lines. As a result, the typical $1/e$ decay time of the sona from the star graph was $\approx1.5\mu s$, a value typical of our 3D experiment.

The perturbed star graph was set up using identical values of parameters as the baseline star graph, except for the length, $L_{n}$. The lengths of the perturbed star graph were chosen to be $PL_{n}$, where $P$ is the perturbation strength. The driving line has zero length in both the baseline and perturbed star graphs.

\subsubsection{Using the Time Domain Approaches \label{sec:3-QuantVolChaStarGraphTD}}

\begin{figure}
\begin{center}
\includegraphics[width=3in]{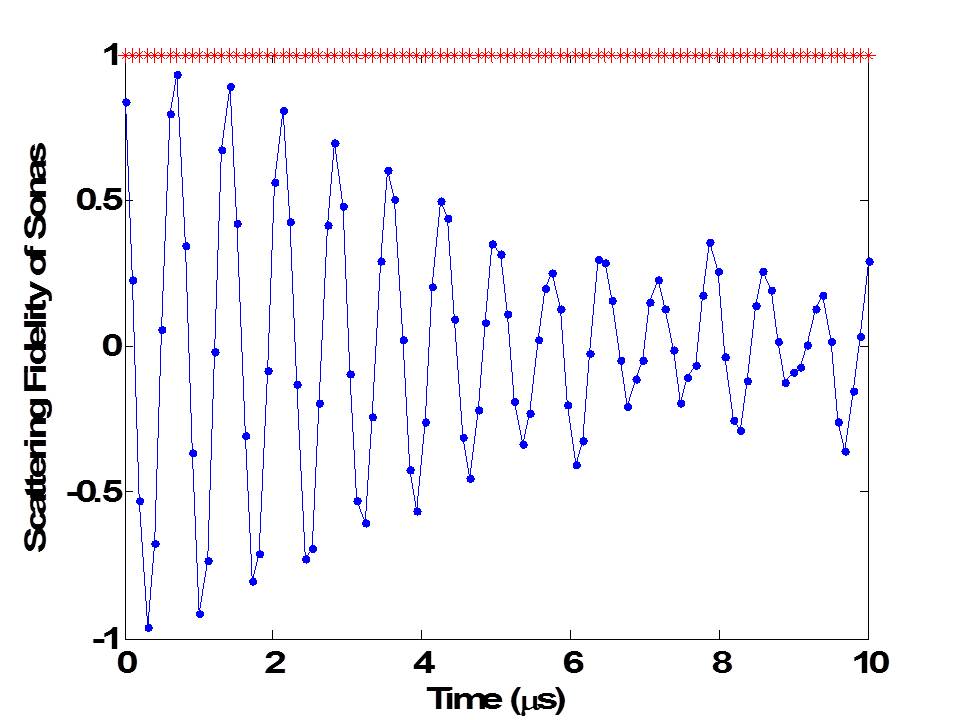}
\end{center}
\begin{quote}
\caption{Scattering fidelity of baseline and perturbed sonas from the star graph before (blue .) and after (red +) optimally scaling the baseline sona along its time axis. The perturbation is a VCSPP with $P=1.0002$. \label{fig:vcpFig10}}
\end{quote}
\end{figure}

A Gaussian pulse of width $1ns$ and center frequency $7GHz$ was used to generate a baseline and perturbed sona from the baseline and perturbed star graphs. The time domain sensing technique which is based on scattering fidelity was applied to quantify a perturbation of strength $P=1.0002$. Fig.~\ref{fig:vcpFig10} shows the scattering fidelity of the baseline and perturbed sonas before (blue) and after (red) optimally stretching the baseline sona. This demonstrates that the scattering fidelity sensing technique can be used to quantify a VCSPP in the quasi-1D chaotic system. The result gives the clearest evidence to the discussion of fidelity decay induced by VCSPPs in Sec.~\ref{sec:3-ResultsSF} (i.e. the results shown in Figs.~\ref{fig:vcpFig5} and \ref{fig:vcpFig6}). VCSPPs induce a scattering fidelity oscillation that is superimposed on a fidelity decay (see Fig.~\ref{fig:vcpFig10}).

\begin{figure}
\begin{center}
\includegraphics[width=3in]{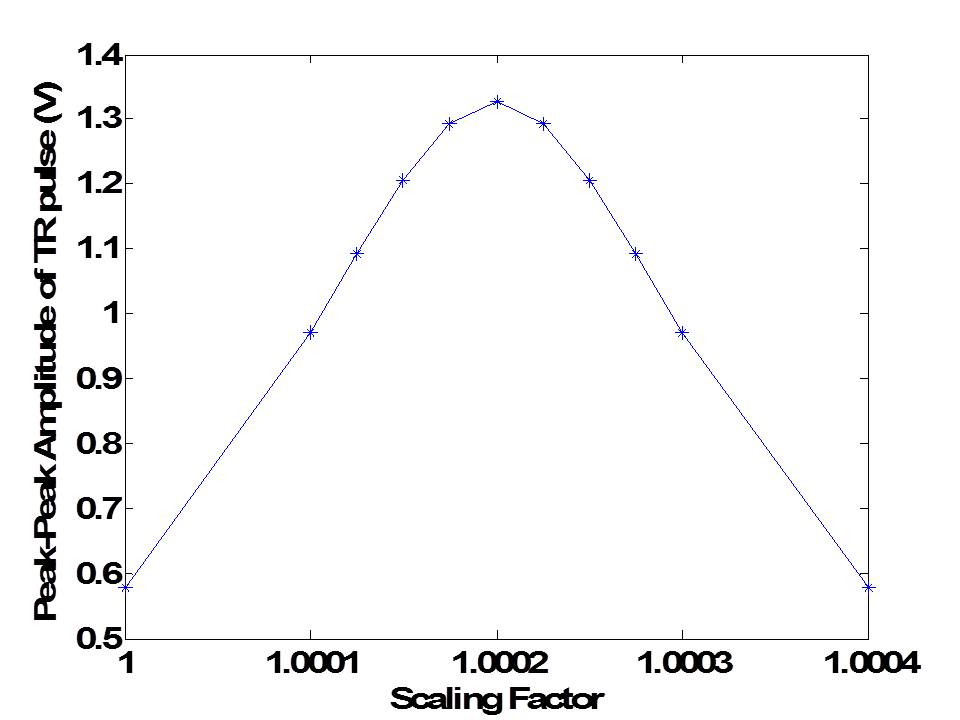}
\end{center}
\begin{quote}
\caption{Peak-to-peak amplitude (PPA) of time reversed pulse reconstructed inside the perturbed star graph using the baseline sona which is scaled along its time axis by different scaling factors. \label{fig:vcpFig11}}
\end{quote}
\end{figure}

The application of the time domain sensing technique which is based on time reversal is presented in Fig.~\ref{fig:vcpFig11}. Fig.~\ref{fig:vcpFig11} shows the PPA of the time reversed pulse reconstructed using the baseline sona scaled with different factors along its time axis. The optimal PPA was obtained when the sona was scaled by a factor exactly equal to the stretching of the transmission line lengths (i.e. $1.0002$).

\begin{figure}
\begin{center}
\includegraphics[width=3in]{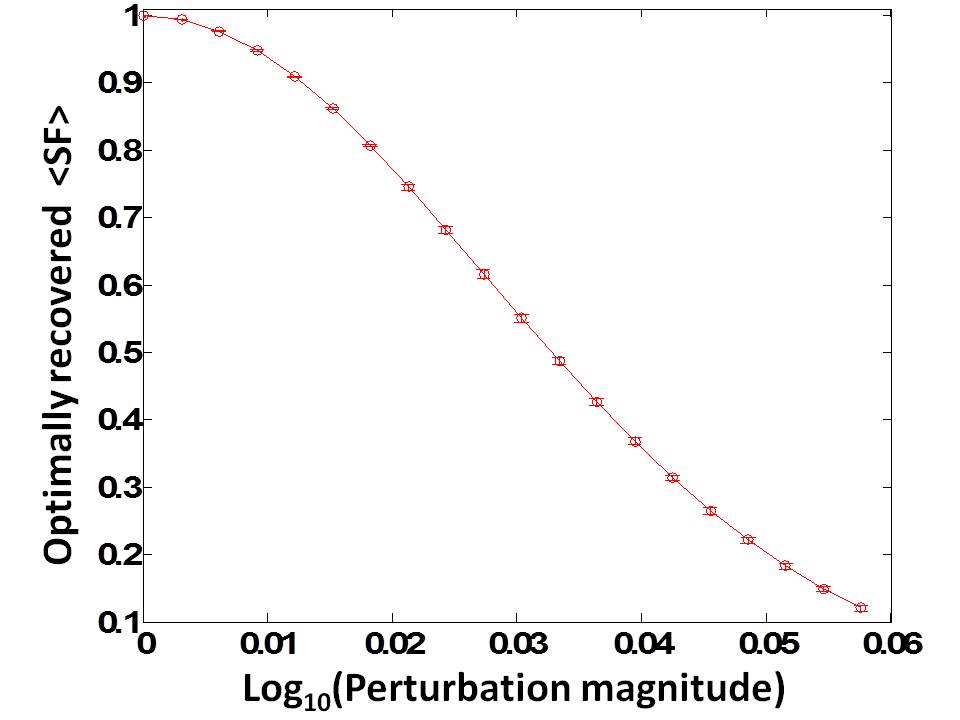}
\end{center}
\begin{quote}
\caption{The average $SF(t)$ of the baseline sona and perturbed sonas, which are optimally scaled by $P$ (perturbation magnitude) along their time axis. The sonas were collected from the baseline star graph, and a perturbed star graph (VCSPP with scaling factor $P$). The average $SF(t)$ was taken from time $t=0s$ to $t=10\mu s$ of the sonas; error bars show the associated standard deviation.  \label{fig:vcpFig12}}
\end{quote}
\end{figure}

The perturbation strength of $P=1.0002$ was shown to be detectable using time domain techniques in the star graph set up described. However, as the perturbation got stronger, the shortcoming of the time domain techniques was revealed. For perturbation strength values, $P$, ranging from $1.0002$ to $1.14$, the $SF(t)$ of the perturbed sona and the optimally scaled (along the time axis) baseline sona was examined. For each $P$, the $SF(t)$ was averaged over time, $t$ (from $0s$ to $10\mu s$, which is the duration of the sonas). The closeness of the average $SF(t)$ to $1$ indicates the success of the time domain technique to undo the effect of the VCSPP, and to measure it. Fig.~\ref{fig:vcpFig12} shows the average value of $SF(t)$ (for optimally scaled baseline sona) versus $\log_{10}(P)$; the standard deviations of $SF(t)$ taken over $t$ are also shown as an error bar in Fig.~\ref{fig:vcpFig12}. As $P$ increased beyond about $10^{.025}\approx1.06$, the effectiveness of the time domain sensing technique starts to deteriorate.

Based on the discussion in Sec.~\ref{sec:3-LimitationTD}, VCSPP perturbations that are characterized by $P > \frac{f_{max}}{f_{min}}\approx\frac{7.5GHz}{6.5GHz}\approx1.15$ are not expected to be quantified using the time domain techniques. The mean spacing between resonant frequencies of the baseline star graph is $\Delta f=\frac{c}{2\sum_{n=1}^{n=N}L_{n}}\approx2MHz$, where $N$ is the number of lines in the star graph \cite{Kottos1999}, and $f_{max}-f_{min}=1GHz \gg \Delta f$. Therefore, the probing pulse excites several resonances of the baseline cavity. The result illustrated in Fig.~\ref{fig:vcpFig12} shows that VCSPP perturbations that are characterized by $P \ll 1.15=10^{0.06}$ are quantifiable.

\subsubsection{Using the Frequency Domain Approach \label{sec:3-QuantVolChaStarGraphFD}}
The limitation of the time domain approach was discussed in Sec.~\ref{sec:3-LimitationTD}, and demonstrated using the star graph in Sec.~\ref{sec:3-QuantVolChaStarGraphTD}. As shown in Fig.~\ref{fig:vcpFig12}, for a strong perturbation such as $P=1.14$, the time domain sensing techniques fail to undo the effect of the VCSPP, and hence to measure it. Here, a frequency domain approach is used to illustrate how the VCSPP with $P=1.14$ can be measured.

\begin{figure}
\begin{center}
\includegraphics[width=3in]{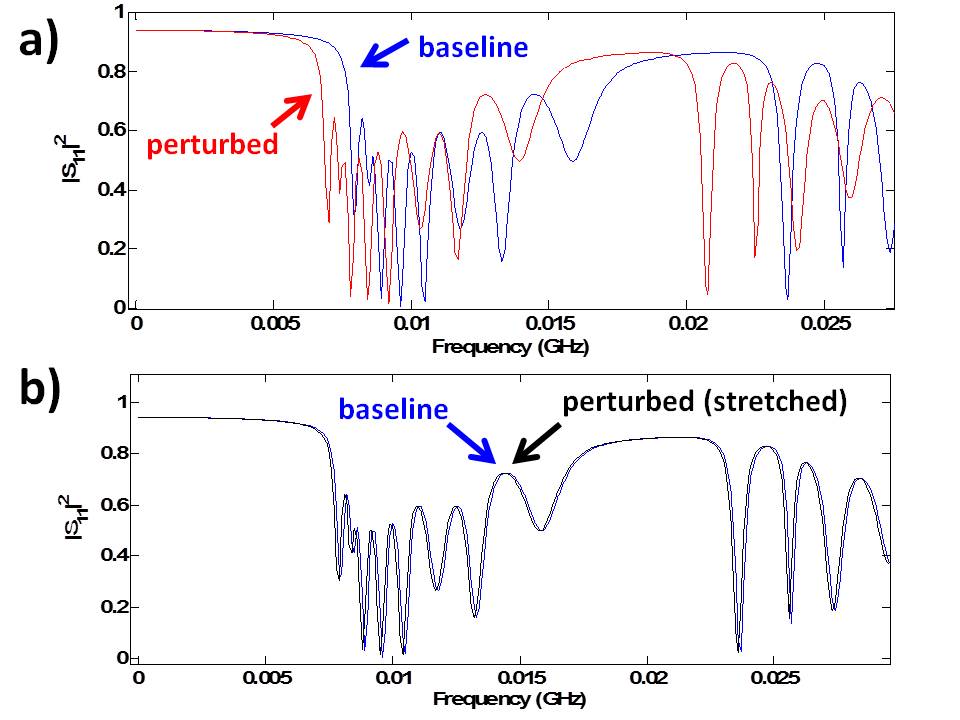}
\end{center}
\begin{quote}
\caption{Quantifying a volume changing perturbation in the frequency domain. (a) The $|S_{11}|^{2}(\omega)$ of the baseline (blue) and a perturbed (red) star graph for perturbation strength $P=1.14$. (b) The $|S_{11}|^{2}(\omega)$ of the baseline star graph (blue), and the $|S_{11}|^{2}(\omega)$ of the perturbed star graph (black) after optimal frequency scaling which measures the VCSPP. \label{fig:vcpFig13}}
\end{quote}
\end{figure}

Fig.~\ref{fig:vcpFig13}(a) shows the $|S_{11}|^{2}(\omega)$ of the baseline (blue) and perturbed (red) star graphs. The $|S_{11}|^{2}(\omega)$ of the perturbed star graph (larger in size by a factor of $P=1.14$) is compressed along the frequency axis compared to the baseline's case. This effect was predicted in Sec.~\ref{sec:3-Theory}, and schematically illustrated in Fig.~\ref{fig:vcpFig1}. The frequency domain approach to measure VCSPP involves optimally scaling the frequency axis of the scattering parameter of the perturbed system (star graph in this case) to align it with the scattering parameter of the baseline system. Fig.~\ref{fig:vcpFig13}(b) shows the $|S_{11}|^{2}(\omega)$ of the baseline star graph (blue) and the optimally stretched $|S_{11}|^{2}(\omega)$ of the perturbed star graph (black). The optimal frequency scaling factor was $P=1.14$, which also successfully measures the VCSPP induced on the baseline star graph. To conclude, given the scattering parameters of a baseline and a perturbed system, one can check if a VCSPP happened, and quantify it.

\section{Discussion \label{sec:3-discussion}}
The development of this quantitative sensor opens up possibilities for several potentially useful applications. For example, when a cavity that has a homogenous material make up is cooled down (or warmed up), it is interesting to check if the temperature stays uniform throughout all parts of the cavity. The sensor developed in this paper would allow one to check if the volume of the cavity is decreasing (or increasing) while the shape is intact; the sensor would also allow one to measure by how much the volume (and hence the temperature) of the cavity is changing. This is feasible as long as the temperature change primarily affects the volume of the cavity, and not the dielectric constant of the medium filling the cavity. This is one possible application of the VCSPP sensor which can practically compete with the traditional option of installing thermometers throughout the cavity. Another possible application of the VCSPP sensor is monitoring if a fluid has displaced another fluid uniformly throughout a cavity. Since the speed of the waves inside different fluids can be different, the displacement of the fluid can change the volume of the cavity, as seen by the waves. This assumes that other wave properties (such as dissipation and dispersion) of the two fluids are similar.

\section{Conclusion \label{sec:3-conclusion}}
Measuring the effects of perturbation to a wave chaotic enclosure and uniquely identifying the perturbation that gave rise to this change is a challenge because many such perturbations can give rise to the same measured change. However, the effect of a perturbation that changes the volume but keeps the shape intact can be theoretically predicted. The theoretical prediction is most clear in the frequency domain. Thus, quantifying volume changing perturbations is best done in the frequency domain. Nonetheless, time domain approaches can be preferred for practical purposes. The time domain approach is limited by a maximum perturbation that can be measured. This limitation was demonstrated using a simulation of a star graph, which is a representative wave chaotic system. The time domain approach can work using either scattering fidelity techniques or time reversal mirrors. Quantification of a volume changing perturbation was experimentally demonstrated using these techniques. The volume changing perturbation was induced experimentally by changing the electrical volume of the cavity. The results of the experiment were compared with FDTD simulation results of the cavity and good agreement was found.

This work is supported by: ONR MURI grant N000140710734; AFOSR grant FA95500710049; ONR/AppEl, Task A2, through grant N000140911190; and the Maryland Center for Nanophysics and Advanced Materials. The computational resources for the FDTD simulations were granted by CINECA (Italian Supercomputing Center) under the Project ISCRA HP10BYQKHN. We would like to thank Jen-Hao Yeh, Matt Frazier, and T.H. Seligman for helpful discussions. Franco Moglie acknowledges the financial support and the hospitality of the Wave Chaos group at IREAP, University of Maryland.


\clearpage

\end{document}